\documentstyle[aps]{revtex}
\begin{document}

\title{Many-Body effects and resonances in universal quantum sticking \\
of cold atoms to surfaces}
\author{Eric. R. Bittner\footnote{email: bittner\@julia.cm.utexas.edu }}
\address{Department of Chemistry and Biochemistry,\\
        The University of Texas at Austin,\\
	Austin, Texas 78712-1167}
\author{John C. Light \footnote{email: light\@hpsi.uchicago.edu}}
\address{The James Franck Institute
and the Department of Chemistry,\\
	The University of Chicago,\\
	Chicago, Illinois 60637}
\date{October 31, 1994}
\maketitle
\begin{abstract}
The role of shape resonances and many-body effects on universal
quantum sticking of ultra cold atoms onto solid surfaces is examined
analytically and computationally using an exactly solvable
representation of the Dyson equation. We derive the self-energy
renormalization of the the transition amplitude between an ultra cold
scattering atom and the bound states on the surface in order to
elucidate the role of virtual phonon exchanges in the limiting
behavior of the sticking probability. We demonstrate that, to first
order in the interactions for finite ranged atom-surface potentials,
virtual phonons can only rescale the strength of the atom-surface
coupling and do not rescale the range of the coupling.  Thus,
universal sticking behaviour at ultra-low energies is to be expected
for all finite ranged potentials.  We demonstrate that the onset of
the universal sticking behavior depends greatly on the position of the
shape resonance of the renormalized potential and for sufficiently low
energy shape resonances, deviations from the universal
$s(E)\propto\sqrt{E}$ can occur near these energies.  We believe that
this accounts for many of the low energy sticking trends observed in
the scattering of sub-millikelvin H atoms from superfluid $^4$He
films.
\end{abstract}

\section{Universal Quantum Sticking}

The exact limiting behavior of the quantum sticking probability has
been repeatedly discussed over the past 25 years, with various authors
arriving at various conclusions
\cite{Goodman71a,Goodman79,Knowles77,brenig80,brenig82,brenig92,Carraro92,Kohn92,Bittner94a}.
Since the limiting behavior is expected only at extremely low
scattering energies, only recently have direct experimental
measurements of sticking at the required energies been obtained, most
notably in the sticking of spin polarized hydrogen to liquid He
films~\cite{Doyle91,Doyle93} and in the desorption of cold positronium
from metal surfaces~\cite{Mills91}.  At source of the theoretical
controversy is the role that the many-body aspects of the scattering
might play at extremely low energies and the effect of the long ranged
behavior of the attractive part of the static potential.

The first complete formal analysis of the problem was by Goodman in
the early 1970's ~\cite{Goodman71a}.  Using a WKB analysis of the low
energy wavefunction, it was demonstrated that that $\lim_{E\rightarrow
0} s(E) > 0 $ for potentials with ranges longer than exponential and
vanishes for those with ranges less than exponential even within the
distorted wave Born approximation (DWBA) and that many-body effects
are not needed in order to account for non-vanishing
sticking~\cite{Goodman79} However, it must be pointed out that these
results are semi-classical and that the WKB approximation used to
derive their sticking results may not permit rigorous extrapolation of
$s(E)$ for energies below the WKB validity condition.  Interestingly
enough for attractive inverse power law potentials with a
characteristic exponent $r \le 2 $, the WKB approximation is valid
down to $E=0$ and classical and quantum predictions agree to within a
constant factor.~\cite{brenig92}

In support of these results Knowles and Suhl~\cite{Knowles77} have
constructed a variational polaron model in which the scattering atom
is dressed by its interaction with the surface phonons.  Within such a
basis, the adatom becomes a quasi-particle with an effective mass
$m^*$ which is position dependent and much larger than the bare mass
$m$ over the potential region.  This leads to an increase in the
density of low energy resonances which increases penetration of the
wavefunction into the potential region and leads to non-vanishing
sticking even for square well potentials. Later, Brenig and
co-workers~\cite{brenig80,brenig82,brenig92} showed that within the
DWBA: (1) sticking should always
vanish for short ranged wells, (2) sticking can be enhanced by
polarization induced resonances for longer ranged potentials, (3) for
the Coulombic potential, quantum and classical analysis yield
identical results.  Recently Carraro and Cole~\cite{Carraro92} applied
Brenig's equations to study H sticking to liquid He film and obtained
remarkable agreement with the experimental results~\cite{Doyle93}.

As far as we know, the most comprehensive examination of the quantum
sticking problem is that of Cloughtery and Kohn who developed an
analytic model for quantum sticking.~\cite{Kohn92} Working up from a
surface composed of N discrete lattice atoms to the continuum,
$N\rightarrow\infty$, they demonstrated that within their one
dimensional model, while polarization effects due to virtual phonon
exchanges may lead to an increase in the sticking coefficient,
eventually these effects will be lost as pure reflection becomes more
important at low energy.

Experimental evidence for quantum sticking has come from a variety of
different sources. Most notably in the desorption of positronium (Ps)
from Al surfaces~\cite{Mills91} and the sticking of spin polarized
hydrogen atoms on liquid $^4$He films~\cite{Doyle91,Doyle93}.  Slow Ps
atoms, composed of an electron and a positron, are almost ideal atoms
for studying quantum sticking. At experimentally accessible scattering
energies of 7.5 meV, the thermal de Broglie wavelength of Ps is on the
order of 100 \AA\ and is much larger than the range of the potential well.
Measurements of the thermal desorption rates of Ps from clean Al(111)
along with detailed balance arguments indicate that the Al surface is
a ``blackbody'' for Ps emission and hence the system fails to exhibit
perfect reflection of ultraslow Ps atoms.  For the case of H on He
films, the interactions are considerably weaker and there are a number
of examples of experiments in which the sticking data extrapolates to
$s(0) = 0$
\cite{H-He1,H-He2,H-He3,H-He4,H-He5,H-He6,H-He7,H-He8,H-He9}.
Surprisingly enough, recent data for H sticking onto thin
He films, reported a gradual {\em increase} in the sticking
probability as the temperature decreased between 10 mK and .1
mK~\cite{Doyle91,Doyle93}. This prompted Hijmans, Walraven, and
Shlyapnokov~\cite{Walraven92} to propose that the atom-surface
potential might be affected by very long-range van der Waals forces
(with relativistic retardation) from the underlying substrate.
Indeed, calculations by Hijmans, Walraven, and
Shlyapnokov~\cite{Walraven92} and by Carraro and Cole~\cite{Carraro92}
indicate that substrate effects could account for this trend and
predict that the $\sqrt{T}$ behavior will be recovered at slightly
lower temperatures.  More recent work which measured the
sticking coefficient as a function of the He film thickness gave
reasonable agreement with the theoretical predictions in temperature
regimes above .1 mK ~\cite{Doyle93} That is, as the film thickness was
increased, the sticking probability was reduced and eventually the
universal $s(T)\propto\sqrt{T}$ behavior was observed.

In this paper we wish to examine two issues which we believe have not
been adequately addressed, and yet are necessary in order to properly
interpret the experimental results mentioned above.  The first issue
is the role of many-body effects on the limiting behavior, and the
second is the role of low energy shape resonances.  In the next
section, we examine the many-body contributions by solving formally
the Lippmann-Schwinger equation for the scattering wavefunction in
which we include the self-energy due to the virtual phonon transitions
between the scattering wave and the inelastic channels.  Since our
theory incorporates the dynamical evolution of the surface directly
into the final equations of motion, it is essentially exact (for
single phonon exchanges) and non-perturbative.  Using approximate
forms of the low energy scattering wavefunction, which we show to
become exact as $k\rightarrow 0$, we derive the renormalization of
the sticking probability due to the the many-body interactions for
realistic atom-surface interactions. We then demonstrate that polarization
due to many-body effects serves only to rescale the strength of the
interactions and does not rescale the effective range.  We next examine
the effect of low lying shape resonances and demonstrate the effect of
increased penetration of the scattering wave into the potential region
near these resonances. Finally, in the last section we apply the
methods which we developed to examine the low energy limiting behavior
of H sticking to thin and thick $^4$He films.

\section{Theory}
The starting point of our theory is the Lippmann-Schwinger (or Dyson)
equation for the scattering wavefunction in which we include
explicitly the self-energy or polarization due to virtual phonon
exchanges.  The self-energy reflects the many-body nature of the
problem and we shall assume that only single exchange processes are
important.  Close coupled forms of this equation have been used
extensively by Stiles, et. al~\cite{Stiles86,Stiles88} and
Jackson
{}~\cite{Jackson86,Jackson88,Jackson89,Jackson90,Jackson91a,Jackson91b,Jackson92}
to study inelastic molecule-surface scattering and Whaley and Bennett
have used the formalism to study atom scattering from disordered
surfaces~\cite{Whaley91}.  We write the elastically scattered
component of the wavefunction as
\begin{eqnarray}
\phi_o(E) = \phi_{bare}(E) + G_{bare}(E) V_{self}(E,T)
\phi_o(E),
\label{eq:ls}
\end{eqnarray}
where $\phi_{bare}(E)$ and $G_{bare}$ are the bare
wavefunction and Green's functions in the absence of surface vibrations and
$V_{self}(E,T)$ is the self energy operator
\begin{eqnarray}
V_{self}(E,T) = {1\over{N}}\sum_q V_q^{ph} (G^+_q(E,T) + G^-_q(E,T))
V_q^{ph}.\label{Self_Energy}
\end{eqnarray}
Our notation is such that
superscripted $+$ and $-$ refer to whether or not a virtual phonon was
initially created or annihilated. Also, throughout this paper we shall
define $H_o$ as the static (zero-phonon) Hamiltonian and $V_q^{ph}(z)$
as the force between the scattering atom and a phonon with wavevector
$q$. We have also assumed that the bath is harmonic.

Closure is accomplished in the usual way by iteratively substituting
$\phi_o$ back into the Lippmann-Schwinger equation and analytically
performing the summation over single phonon exchange diagrams.
\begin{eqnarray}
\phi_o &=& \phi_{bare} +  G_{bare}V_{self}\phi_{bare} + \cdots
\nonumber \\
&=& \left(1 - G_{bare}V_{self}\right)^{-1}\phi_{bare}
\end{eqnarray}

Before applying our theory to a physically realistic system, it is
important to explore the effects induced by the additional
polarization due to virtual phonon exchanges.  Let us consider the
``golden rule'' transition rate between a very low energy scattering
state and a single bound state, $\psi_B$, and attempt to relate the
transition probabilities predicted using a dressed wave to those
predicted from using the bare wavefunction.  In short, we need to
consider the matrix elements
\begin{eqnarray}
\langle \psi_B | V^{ph} | \phi_o\rangle  =
\langle\psi_B|
 \left\{ 1 - V^{ph} G_{o}(E) V^{ph}(G^+(E) + G^-(E))\right\}^{-1}
V^{ph}|\phi_{bare}\rangle,\label{eq:GR}
\end{eqnarray}
where $G_o(E)$ is a solution of
\begin{eqnarray}
(H_o - E)G_o(z,z') = \delta(z-z')
\end{eqnarray}
subject to the boundary condition $G_o(-\infty,z) = 0$ and
$G_o(z,z')\rightarrow 0 $ as $z\rightarrow\infty$.

To evaluate this matrix element we need to make a series of
approximations regarding the functional form of $\phi_{bare}$.  The
bare scattering wave, $\phi_{bare}$, is an solution of
$(H_o-E)\phi_{bare}=0$ subject to the boundary conditions:
$\phi_{bare}(-\infty) = 0$ and $\phi_{bare}(z)\rightarrow \sin(k z) +
\tan(\delta_k)\cos(kz)$ as $z\rightarrow+\infty$.  We can also define
the real regular and irregular functions, $\psi_{r,i}(z)$, as
solutions of
\begin{eqnarray}
(H_o - E) \psi_{r,i}(z;E) = 0
\end{eqnarray}
subject to the boundary conditions
\begin{eqnarray}
\psi_r(z,k)&\stackrel{z\rightarrow\infty}{\longrightarrow}& \sin(kz +
\delta) \nonumber \\
\psi_i(z,k)&\stackrel{z\rightarrow\infty}{\longrightarrow}& \cos(kz +
\delta). \nonumber \\
\end{eqnarray}
Using these functions and the Wronskien relation, we can write
$G_o(z,z';k)$ as
\begin{eqnarray}
G_o(z,z';k) = -{1\over{k}}\left\{
\begin{array}{ll}
\psi_r(z;k)\psi_i(z';k) + i \psi_r(z;k)\psi_i(z';k) & z \le z' \\
\psi_r(z';k)\psi_i(z;k) + i \psi_r(z;k)\psi_i(z';k) & z > z'
\end{array}\right. ,
\end{eqnarray}
where $k=\sqrt{2mE}$ is the scattering momentum.

For very low energies and temperatures, virtual transitions are
predominantly to the bound state and we can obtain the single phonon
exchange propagator from the bound (asymptotically closed)
wavefunctions, weighted by the appropriate phonon densities of states,
which in the continuum limit can be taken as
\begin{eqnarray}
G^+(z,z') & = &
\psi_B(z)\psi_B(z')
\int\frac{d^2q}{2\pi}
\frac{\cos^2(qa/2)}{\omega(q)(E_B-\omega(q)+k^2)} (n_q(T) + 1)\nonumber \\
&=& \psi_B(z)\psi_B(z')\int_{0}^{\omega_D}
d\omega\rho(\omega)
\frac{\cos^2\left(q(\omega)a/2\right)}{(k^2+E_B-\omega+i\eta)\omega}
(n(\omega,T)+ 1).
\end{eqnarray}
where $\omega_D$ is the Debye frequency of the lattice, $n_q(T)$ and
$n(\omega,T)$ are the Bose-Einstein occupation numbers, and
$\rho(\omega)$ is the density of states.  The integral can be
evaluated by choosing the appropriate contour which avoids the pole at
$\omega = k^2+E_B$~\cite{Kohn92} and the result is (using a Debye
model for the phonons)
\begin{eqnarray}
G_k^+(z,z') &=& \psi_B(z)\psi_B(z'){\cal F}(k),
\end{eqnarray}
where
\begin{eqnarray}
{\cal F}(k) &=&{2\over{\pi\omega_D^2}}
-\!\!\!\!\!\!\int^{\omega_D}_{0} {1\over{k^2+E_B-\omega}}(n(\omega) + 1)
\left[{\omega_D^2-\omega^2\over{\omega^2}}\right]^{1/2}
d\omega\nonumber\\
&-& i {2\over{\omega_D^2}}\left({\omega_D^2-(k^2 + E_B)^2\over{(k^2 +
E_B)^2}}\right)^{1/2}.
\end{eqnarray}
The notation $-\!\!\!\!\!\int^{a}_{b} dx$ means that we take the
Cauchy principal value when evaluating the integral.

Putting all this together yields an expression for the matrix element
which is convenient to evaluate
\begin{eqnarray}
\langle\psi_B|V^{ph}|\phi_o\rangle = \int dz \int dz'
 \psi_B(z)\langle z
| 1 - V^{ph}G_oV^{ph}G^+_k|z'\rangle^{-1}V^{ph}(z')\phi_{bare}(z').
\end{eqnarray}
Here, self energy matrix element is given by
\begin{eqnarray}
\langle z| 1 - V^{ph}G_oV^{ph}G^+_k|z'\rangle =
\delta(z-z')-V^{ph}(z)\psi_B(z'){\cal F}(k) I(z;k),
\end{eqnarray}
where
\begin{eqnarray}
I(z;k) =
\left[\int dz''
G_o(z,z'';k)V^{ph}(z'')\psi_B(z'')\right].
\end{eqnarray}
 In order to take the $k\rightarrow 0$ limit of these expressions, we
first need to make an assumption regarding the extreme low energy
behavior of the $\psi_{0,1}(z;k)$ functions used to construct $G_o$.
(Justification for this assumption for the $1/z^3$ potential is
included in the Appendix.)
\begin{eqnarray}
\lim_{k\rightarrow 0}\psi_r(z;k) &\approx & k \psi_r(z,k=0) \\
\lim_{k\rightarrow 0}\psi_i(z;k) & \approx & \psi_i(z,k=0)
\end{eqnarray}
where $\psi_{r,i}(z,k=0)$ are solutions of $H_o \psi_{0,1}(z) = 0$
which are regular and irregular at the surface and are {\em
independent} of $k$.  Using these, we can construct the $E\rightarrow
0$ Green's function
\begin{eqnarray}
\lim_{k\rightarrow 0}G_o(z,z';k) =
\left\{
\begin{array}{ll}
\psi_r(z;k=0)\psi_i(z';k=0) + i k\psi_r(z;k=0)\psi_r(z';k=0) & z \le z' \\
\psi_r(z';k=0)\psi_i(z;k=0) + i k\psi_r(z;k=0)\psi_r(z';k=0) & z > z'
\end{array}
\right.
\end{eqnarray}
Thus the integral, $I(z;k)$, which occurs in the self-energy
expression can be broken into real and imaginary components, both
independent of $k$.  That is
\begin{eqnarray}
\langle z| 1 - V^{ph}G_oV^{ph}G^+_k|z'\rangle
\stackrel{k\rightarrow 0}{=} \delta(z-z')
- V^{ph}(z)\psi_B(z')(I_1(z) + ik I_2(z)){\cal F}(k),
\end{eqnarray}
where
\begin{eqnarray}
I_1(z) &=& \Re e I(z;k\rightarrow 0) \\
I_2(z) &=& \Im m I(z;k\rightarrow 0).
\end{eqnarray}

Finally, substituting this back into Eq.~\ref{eq:GR} yields for small
$k$
\begin{eqnarray}
\langle\psi_B|V^{ph}|\phi_o\rangle &=&
\langle\psi_B|V^{ph}|\phi_{bare}\rangle ( 1 + \Sigma(k) + \Sigma^2(k)
+ \cdots )\nonumber \\
&=&\langle\psi_B|V^{ph}|\phi_{bare}\rangle /( 1 - \Sigma(k)),
\label{eq:renorm}
\end{eqnarray}
where $\Sigma(k)$ is the polarization renormalization due to virtual
phonon exchanges ( in the small $k$ limit).
\begin{eqnarray}
\Sigma(k) = {\cal F}(k)\int dz V^{ph}(z)(I_1(z) + ik I_2(z)).
\end{eqnarray}
Since the integral in $\Sigma(k)$ is at most linearly dependent on
$k$, any deviations from the DWBA results are likely to come from
${\cal F}(k)$.   We note that the denominator, \(1-\Sigma(k)\),
should approach a non-zero limit as $k\rightarrow 0$. The limiting low
energy behaviour will thus be determined by the perturbation integral,
\(\langle \psi_B|V^{ph}|\phi_{bare}\rangle\), which approaches zero as
$\sqrt{E}$ in the limit of $k\rightarrow 0$ for finite ranged
atom-phonon interactions.

What we have neglected in this evaluation is the possible contribution
from shape and threshold resonances. In particular, a threshold
resonance or a shape resonance nearly at the threshold may produce an
additional pole close enough to the real energy axis to change the
limiting behavior of ${\cal F}(k)$.  For the moment, let us consider
the case that there are no threshold resonances and that we are well
below the last shape resonance of the static potential surface.  In
this case, $\Sigma(0)$ is a constant between 1 and -1 and
Eq.~\ref{eq:renorm} produces only a rescaling correction to the DWBA
result, i.e. there is no change in the effective range of the
potential and the $s\rightarrow\sqrt{E}$ limiting behaviour should hold.

Interestingly enough, the calculations which we presented in
Ref.~\cite{Bittner94a}, did not show the limiting behavior predicted
above and we have identified the reasons.  In that paper, we computed
the sticking probability as a function of surface temperature at a
scattering energy $E \rightarrow 0$ for a model He atom scattering
from a featureless Cu surface.  We first tested the model by comparing
finite $E$ results to results published previously for the same model
and obtained good agreement.  We then varied both the attractive range
of the potential, from long ranged $V_{att}\propto -1/z^3$ to the
short ranged $V_{att}\propto -e^{-az}$ and the atom-phonon interaction
by varying the phonon force constants and by changing the effective
mass of the surface atoms.  We then reported that the model produces
sticking as $E\rightarrow 0$ which decreases as the range of the
potential is decreased and decreases as the phonon force constants are
increased.  Since this is precisely the behavior one would expect for
potentials with longer ranged interactions or if the effective
scattering mass of the atom is increased, we concluded that the
inclusion of the virtual phonon processes causes a renormalization of
the range of the static potential and/or an enhancement of the
effective mass of the scattering atom and, hence, leads to
non-vanishing sticking.

Later analysis indicates that these calculations may be in error due
to our particular choice of a linear trial function used in our
numerical calculations to generate the $E\rightarrow 0$ solutions.  In
our original calculations, we tested the method against results
published by Jackson for He scattering from Cu(100) at 5.1 meV and
normal incidence and found satisfactory agreement and convergence. The
problem with the low $E$ results occurs because the finite ranged
basis and the linear auxiliary trial function used imposes an
artificial energy {\em minimum} in the convergence properties of the
equations and the method can not converge wavefunctions with energies
below that critical value.  Usually, convergence errors occur at the
high end of the energy spectrum where one does not have enough high
energy components in the basis used to represent the Green's function
and thus to adequately represent the wavefunction. In the present
calculations, however, we use $\sin(kz)$ as the auxiliary trial
function and do obtain the correct limiting behaviors predicted by the
above analytical model.

In the next section, we apply the equations derived above to study H
reflection from He films.  Since hydrogen is an extremely light atom
and the He-H potential is well characterized this is an ideal physical
system for this type of study.  Also, it has been demonstrated
recently, that the sticking behavior is strongly dependent upon the
long ranged interactions with the underlying
substrate~\cite{Doyle93,Carraro92,Walraven92}.  We shall demonstrate
that the increased polarization from the underlying substrate causes
an increase in the density of low lying shape resonances which enhance
penetration of the wavefunction into the potential region and enhance
sticking.  To elucidate the effect of resonances in the sticking, we
examine sticking on an even more idealized situation in which the H-He
interaction is reduced to that of a square well and H-ripplon
interactions decay exponentially in z.  By adjusting the ``range'' of
the well (at constant well depth) we can very easily see the
enhancement of sticking due to low lying resonances.  In contrast to
our earlier results, once the scattering energy is sufficiently below
the resonances, penetration into the potential region becomes greatly
diminished as reflection from the long ranged part of the potential
begins to dominate and the universal $s(E)\propto \sqrt{E}$ behavior
of the sticking coefficient is recovered.

\section{Sticking of atomic Hydrogen  on liquid Helium films}

The interactions between H and liquid He are known rather well. In
fact, variational estimates of the binding energy of H on liquid
$^3$He and $^4$He date back to the late 70's~\cite{H-He7,H-He1}.  At
long range, the potential can be obtained from summing over all
pairwise van der Waals interactions between the H and the He atoms.
At large distances (over 200\AA), retardation effects need to be
included in order to provide the proper cut off in the attractive
forces.  At shorter ranges, the potential can obtained from empirical
fits to atomic beam scattering data.~\cite{H-He7}.  To lowest order,
the effective H--liquid He potential can be constructed by integrating
over H-He pairwise potentials and assuming that the liquid He is a
semi--infinite and incompressible fluid extending from $z=-\infty$ to
$z = 0$ and with a sharp density profile, $\rho(R,z) =
\rho_o\Theta(z+u(R))$.
\begin{eqnarray}
V(z) = \rho_o\int d^2R'\int_{-\infty}^{+\infty}dz'v_{pair}(r)
\Theta(z'+R') \label{Veff},
\end{eqnarray}
where $r = \sqrt{R^2 + (z-z')^2}$, $v_{pair}$ is the H-He pair
potential, and $\rho_o$ is the number density of bulk liquid $^4$He.
($\rho = 0.0218 $\AA$^{-1}$.)

There have been numerous theoretical treatments of the H-He pair
potential, most recently by Bhattacharya and Anderson using a quantum
Monte Carlo method and the reader is referred to their paper for a
comprehensive listing of previous treatments~\cite{An94}. A
number of forms for the effective potential have been proposed ranging
from Morse potentials used by Zimmerman and Berlinski~\cite{H-He1} to
much more realistic forms which include the $1/z^3$ van der Waals
contributions both from the He film and the underlying substrate.  The
effective potential used in our study is similar to that used by
Carraro and Cole~\cite{Carraro92} and by others~\cite{Goldman}.  We
included both He film and substrate van der Waals polarization
contributions as well as a relativistic cut off factor which limits
the long range behavior of the potential at distances greater than
$\lambda = 200$\AA.
\begin{eqnarray}
V_o = \Delta(z)  - C_3 \left( { z^3\over{(z-z_o)^6  + z_o^6}} -
 {1\over{(z+d)^3}} \right)\gamma(z)
- C_s{1\over{(z+d)^3}}\gamma(z+d),
\end{eqnarray}
where $\Delta(z) = D \exp (-\beta (z-z_o))$ and arises from atomic
core repulsions.  As discussed in the introduction, the final results
should not be very sensitive to the exact form of this part of the
potential.  Retardation effects in the attractive terms are included
in $\gamma(z)$ which we assume to take the form $\gamma(z) =
(1+z/\lambda)^{-1}$ where $\lambda = 200$\AA~ is the characteristic
range of the effect.  The remaining parameters, $\beta$, $z_o$ and
$D$ are tabulated in Table~\ref{tab:params} and were constrained to
produce a single bound state with a binding energy within the range of
the latest experimental estimates of 1.0 $\pm$ .1 K for an infinitely
thick He film.  The polarization terms are $C_3 = 219.7$ K \AA$^3$ and
$C_s = 5000$ K \AA${^3}$.  Finally, we note that any effective
potential, $V$, can be generated by an effective two-body potential,
$v$ by directly inverting Eq.~\ref{Veff}.~\cite{Goldman}
\begin{eqnarray}
v(r) = {1\over{2\pi\rho_o}}{\partial^2\over{\partial z^2}}V(z).
\end{eqnarray}

We treated the He film as a smooth structureless surface.  In this
system, the inelastic effects are mediated through the capillary
motion of the film.  For the temperature ranges in which we are
interested, these capillary waves, called ripplons, provide the only
remaining inelastic channels, the other inelastic channels, such as
those mediated by the phonons and the rotons are effectively frozen
out.  Mathematically, the ripplons and phonons are essentially
equivalent quantities, differing only in dispersion and coupling.  The
displacement of the He surface can be written in terms of the ripplon
operators as
\begin{eqnarray}
u(R) = {1\over{S^{1/2}}}\sum_q \left(
{\hbar q\over{M\rho\omega_q}}\right)^{1/2}
(a_q^\dagger + a_{-q})e^{iq\cdot R}.
\end{eqnarray}
The ripplon dispersion for thick He films is given by the well know
form for surface capillary waves
\begin{eqnarray}
\omega_q^2 = g q + {\alpha\over{M\rho}}q^3,
\end{eqnarray}
where $\alpha$ is the He surface tension ($\alpha =  0.27$ K\AA$^{-2}$)
{}~\cite{Landau:Fluids}, and the ``gravitational'' acceleration is given
by, $g = C_s/d^4$.

As with He scattering from metal surfaces, we make the {\em ansatz}
that the complicated dependence upon $q$ in the atom-ripplon
interaction can be factored into  two terms, one which includes all
information regarding momentum transfer and one which depends upon
only the core interactions at the surface. The  form which we use is
\begin{eqnarray}
V_q^{ph} = \sqrt{q\over{2M\omega\rho}}{2\beta\over{4\beta^2 + q^2}}
\Delta(z).
\end{eqnarray}
The fact that we are using an approximate form for the coupling may
introduce an overall error in our final results; however, the overall
limiting behavior (i.e. whether or not the sticking vanishes)
is expected to be universal and should not depend
strongly upon the exact form of this coupling.

Finally, the sticking probability  is obtained from the transition
probability between the dressed scattering state and the bound states.
\begin{eqnarray}
%s(T)& =& 2 \pi\sum_n \int{d^2q\over{2\pi}}{2m\over{\hbar
%k_z}}|\langle\phi_n|V^{ph}_q|\phi_o\rangle|^2 \nonumber\\
%&\times&\left(
%\right)
s(T,E) ={4 \pi^2 m\over{k_z}} \sum_n
|\langle\phi_n|V^{ph}_q|\phi_o\rangle|^2
\psi''(E/\hbar-\omega_q) /(1-e^{-\hbar\omega_q\beta})
\end{eqnarray}
where $\psi''(\omega)$ is the imaginary part of the ripplon
susceptibility, which can be approximated using a Debye model,
\begin{eqnarray}
\psi''(\omega) = {3 \pi\over{2 M\omega_d^3}}\omega\Theta(\omega_D -
|\omega|),
\end{eqnarray}
where $\omega_D$ is the surface Debye frequency, which we take to be
on the order of 1 meV, or about 12 K.  This produces a cut off in the
ripplon momentum dispersion of $q_D$ = 1 \AA$^{-1}$.

Figure~\ref{fig:hhe1} shows the variation of $s(T)$ for a 30 \AA\ He
film and bulk He as the H scattering energy is reduced below 10 mK.
Superimposed are the experimental data  for the same temperature
regime.  Given the uncertainties in both the experimental data and in
the sensitivity of our calculation to variations in the potential
parameters, the agreement is somewhat satisfactory.  For scattering
from bulk He films, the substrate polarization is attenuated by the
thickness of the film. Although the binding energy of H on bulk He is
somewhat less than that for thin films, the shape resonances are far
enough into the continuum that penetration into the potential region
is not enhanced below 1 mK and quantum reflection begins to dominate.
For thin films, the situation is just the opposite. The additional
polarization due to the underlying substrate dominates the long ranged
behavior of the potential and the density of the shape resonances near
threshold is increased. This gives rise to the observed increase in
the sticking as the film thickness is decreased.  Below about  0.1 mK,
which is at the end of the most recent available sticking data,
barring the presence of a lower lying shape resonance, the resonance
enhancement effect is diminished and quantum reflection should be
observed.

In the next figure, (Figure~\ref{fig:hhe2}), we show the changes in
$s(T)$ as the He film thickness is increased from thin film to bulk
at a constant scattering energy of .35 mK. Again, although our
results differ from the experimental values by perhaps a factor of
two or three on the average, the agreement between our calculations
and the experimental data is on the whole quite good, especially
given the apparent sensitivity of $s(T)$ to the long ranged part of
the potential and the lack of an accurate measurement of the
substrate interactions.  What is striking about our calculations is
the bump in our computed $s(T)$ which appears when the film thickness
is just below 50 \AA.  This is due to the appearance of a threshold
resonance, which eventually changes to a second bound state when the
film thickness is around 15-17 \AA.  In Table 1 we show the variation
in the binding energy as the film thickness is decreased as computed
by diagonalizing the static surface Hamiltonian. Quantitative
comparison to the the experimental data much below this region is
difficult due to other substrate effects which might come
into play, such as roughening, corrugation effects, and inelastic
coupling to the substrate phonon modes.  In fact, the experiments
were done using either sintered silver or epoxy as a substrate. For
the sintered silver case, for coverages above 30 \AA, the pores in
the sinter should be filled~\cite{Doyle93}; however, for coverages
below that the pores may be only partially filled.  In any case, the
experimental data also indicates an abrupt change in the sticking
behavior which occurs when the He film thickness is below 20 to
30\AA.

As a final demonstration of the role of shape resonances in quantum
sticking, we consider the case of H sticking to ``pseudo'' liquid He
where we have replaced the realistic effective potential used above
with a square well.  By adjusting the width of this well, $\lambda$,
we can very easily examine (analytically, if we want), the
enhancements to the sticking due to low lying resonances.  In
Fig.~\ref{fig:square_well} we plot $s(T)$ vs $\lambda$ at constant
scattering energy well into the quantum reflection regime ($E=0.01$
meV). For the narrowest well ($\lambda = 10$ \AA),
the potential barely supports a single bound state with a binding
energy of 0.031558 K.  As $\lambda$ is increased to 50\AA, more bound
states are added to the well.  This affects the sticking in a profound
way.  At each peak in this figure, an additional bound state is added
to the well.  As $\lambda$ increases further, the binding energy of the new
state increases and sticking decreases.

Lastly, in Fig.~\ref{fig:sq2} we consider a rather exceptional case in
which the effective range of the atom-phonon interaction was chosen to
be longer than the effective range of the attractive part of the
potential. Here, we set $1/\beta = 20$\AA\ and vary both the range of
the square well, $\lambda$, and the scattering energy.  Two features
are immediatly appearent. First is that the universal sticking
behaviour is observed in all cases indicating that for finite ranged
atom-phonon interactions, universal sticking is to be expected.
Secondly, is the dramatic enhancement of the sticking due to the low
lying resonances.

\section{Discussion}
We have presented a theory for the sticking of ultra-cold atoms from
surfaces, and we have presented evidence for what we believe to be the
proper limiting behavior of $s(T)$ for non-Coulombic
potentials. Although our model calculations are limited to low
scattering energy where only transitions to bound states are allowed,
we believe that the limiting behavior shown here is in fact correct,
agreeing with the predicted behaviors.  Our calculations indicate that
$s(T)$ can have non-vanishing limiting behavior only in the presence
of extremely low lying (threshold) resonances.  We demonstrate that
many body effects do persist into the quantum sticking regime;
however, their net effect at very low energy is to rescale the
strength of the potential but not to rescale the effective range.
Since sticking is most sensitive to the range of the potential, the
limiting low E behavior becomes independent of the strength of the
interaction and the coupling to the inelastic channels.
At very low scattering energy, there is a competition between
polarization effects due to the virtual phonon exchanges and quantum
mechanical reflection by the long ranged part of the potential.  In
the absence of shape and threshold resonances, quantum reflection is
expected to win out and the universal $s(E)\propto\sqrt{E}$ limit is
obtained.

There are a number of fundamental questions which remain unresolved
both experimentally as well as theoretically.  When the inelastic
interactions are very weak, as in the case of H on He films, subtle
details, such as relativistic retardation effects and substrate
contributions to the potential become very important and can
dramatically change the sticking behavior.  These subtle contributions
are very difficult to quantify experimentally due in part to the
technical difficulties of working in the submillikelvin
regime~\cite{Doyle91}.  However, given the sensitivity of the sticking
to very subtle details of the long range part of the potential, these
low energy experiments offer unique insight to substrate effects and
relativistic retardation effects that would otherwise be unmeasurable.

\section*{acknowledgments}
We wish to thank Professor Steven J. Sibener and Dr. Carl Williams for
for many discussions and suggestions this work.  This work was supported
by the Materials Research Center of the University of Chicago  under
the NSF  grant DMR-88-19860.

\appendix
\section{Zero energy wavefunctions for $1/z^3$ potentials}
The essence of the theory of quantum sticking at very low scattering
energies boils down to the fact that near the classical turning point,
the scattering wavefunction vanishes linearly with momentum, $k$.
Thus, any matrix element coupling this low energy scattering state to
 a bound state will also vanish. A demonstration of this can be seen by
 solving the following Schr\"odinger equation, written in
 dimensionalless units.
 \begin{eqnarray}
 \psi_k'' + (g^2/z^3 + k^2) \psi_k = 0
 \end{eqnarray}
 on $z \ge 1$ and subject to the the boundary condition, $\psi_k(1) =
 0$. Exact analytic solutions for this exist only for $k=0$ and the
 result is~\cite{brenig82}
 \begin{eqnarray}
 \psi_0(z) = A\sqrt{z}\left[N_1(x) - N_1(2g)/J_1(2g) J_1(x)\right],
 \end{eqnarray}
 where $N_1(x)$ and $J_1(x)$ are Neumann and Bessel functions
 with arguments, $x = 2g/\sqrt(z)$.  The asymptotic expansion of
 $\psi_0$ is obtained by taking  $x\rightarrow 0$ and using the known
 asymptotic expansions for the Bessel and Neumann functions
 \begin{eqnarray}
 J_1(x) &=& {x\over{2}} - {x^3\over{16}} + \cdots \\
 N_1(x) &=& {1\over{\pi}}\left[ x \ln (\gamma x/2) - {2\over{x}} +
 {x\over{2}}\right] + \cdots,
 \end{eqnarray}
 where $\gamma$ is the Euler constant \(( \ln \gamma = .5772\cdots)\).
 Also, we can use the $x\rightarrow \infty$ asymptotic forms of $J_1$
 and $N_1$ to obtain
 \begin{eqnarray}
 N_1(2g)/J_1(2g) = \tan(2g - {3\over{4}}\pi).
 \end{eqnarray}
 Putting things together, one obtains
 \begin{eqnarray}
 \psi_0(z) = -A g \left\{\left[ \tan\phi_o +
 {1\over{\pi}}\ln\left({z\over{\gamma^2 g^2}}\right) + {z\over{g^2}} +
 1 \right] + \cdots \right\}.
 \end{eqnarray}

 For $k\ne 0$, approximate solutions can be obtained from the $k=0$
 solution by partitioning the potential into interior and exterior
 portions.  The interior part, \begin{eqnarray} v_o(z) = \left\{
 \begin{array}{ll} g^2/z^3 - k^2 & {\rm for } z \le z_c \nonumber \\ 0
 & {\rm otherwise,} \\ \end{array}\right.  \end{eqnarray} where the
 critical distance, $z_c$ is determined by $k^2 = g/z^3_c$.  From this
 particular choice, it is obvious that for $z \le z_c$, the above
 expression for $\psi_0$ is the solution, and for $z \ge z_c$, the
 asymptotic form, \begin{eqnarray} \psi_k = \sin(k z_c + \delta_k),
 \end{eqnarray} is the solution.  The phase shift, $\delta_k$, and
 coefficient, $A$, are determined by requiring the the logderivative,
 $\gamma_k(z)= \psi_k'/\psi_k$ be continuous at $z = z_c$.  From the
 matching condition, as $k\rightarrow 0$, \begin{eqnarray} A = - \pi g
 k + {\cal O}[k^{5/2}\ln k].  \end{eqnarray} Similarly the phase shift
 can be estimated.  The important observation is that the amplitude of
 the wavefunction inside the interaction region, as defined by $z <
 z_c$, vanishes linearly with $k$.  This is to say, that there is no
 transmission of probability flux past $z_c$, and thus the
 wavefunction is totally reflected by the attractive part of the
 potential before it even reaches the spatial region in which it can
 interact strongly enough with the phonons to cause sticking to occur.
 This is the origin of universal quantum reflection.

\vspace{2.00in}
\begin{table}[h]
\caption{Effective Potential Parameters and Computed Binding Energies
for H on Bulk $^4$He and on thin $^4$He films ($d$ = 30 \AA). The
reported value of $E_o$ for bulk $^4$He is 1.0K (See
Ref.~\protect{\onlinecite{Doyle93}}.)}\label{tab:params}

\begin{tabular}{ccccc}\hline
$D$ [K ]  &   $\beta$ [\AA$^{-1}$] & $z_o$ [\AA ]   & $E_o$ [K](thin)& $E_o$
[K] (bulk) \\
\hline \\
3.0          &    0.20                &  4.8           & -0.981589        &
-0.855165 \\
2.9          &    0.20                &  4.8           & -1.059139        &
-0.931325 \\
2.8          &    0.19                &  4.8           & -1.155515        &
-1.025170 \\
2.8          &    0.20                &  4.8           & -1.139079        &
-1.009945 \\
2.8          &    0.21                &  4.8           & -1.122955        &
-0.994995 \\
2.7          &    0.20                &  4.8           & -1.221373        &
-1.090980 \\
2.6          &    0.20                &  4.8           & -1.305987        &
-1.174390 \\
\hline
\end{tabular}
\end{table}

\begin{figure}[h]
\caption{Trapping  probability versus atom energy for H onto a 3 nm
thick and bulk He films (1 mm).  The curves labeled A and B are the
results obtained from our calculations for thin and thick He films.
Superimposed are the experimental results for each of these limits.}
\label{fig:hhe1}
\end{figure}

\begin{figure}[h]
\caption{Sticking probability versus He film thickness.  In each case,
the scattering energy and surface temperature was .35 mK. The diamonds
are the experimental data points from Ref.[\protect{\cite{Doyle91}}] The bump
in our
computed data is due to a threshold resonance which appears when the
film is less than about 30 \AA\ thick. }\label{fig:hhe2}
\end{figure}

\begin{figure}[h]
\caption{Sticking  vs. well width for a square well potential at
constant scattering energy. (well depth = 2.8 K) The increases in the
sticking are due to the the appearence of additional weakly bound states
as the well width is increased. For the narrowest case considered, the
well supports a single bound state almost at threshold (binding energy
= 0.03 K). Threshold well widths for each additional bound state are
indicated by arrows.}
\label{fig:square_well}
\end{figure}

\begin{figure}[h]
\caption{Sticking vs. Scattering energy for a square well potential.
Here we have chosen the effective range of the atom-phonon to be 20\AA\
and vary the width parameter, $\lambda$ from 9.5\AA\ to 21.5\AA\ in
which we pass from 1 bound state to 2 bound states in the square well
potential.  See text for details.}
\label{fig:sq2}
\end{figure}

%%
%% These figures can be obtained via. anonymous ftp to
%% jcp.uchicago.edu (jcp express) under /pub/009506
%%
%\epsfbox{Fig1.ps}
%\epsfbox{Fig2.ps}
%\epsfbox{Fig3.ps}
%\epsfbox{Fig4.ps}

\end{document}